\documentclass[12pt,prl,aps,superscriptaddress]{revtex4}
\usepackage{graphicx}
\usepackage{amssymb,bm}
\newcommand{\beq}[2]{\begin{equation}#1\label{#2}\end{equation}}
\newcommand{\ceq}[1]{(\ref{#1})}

\newfont{\mbld}{cmbx10 scaled 800}
\newfont{\cab}{cmsy10 scaled 1200}
\newfont{\scab}{cmsy10 scaled 1000}
\newfont{\bcall}{cmbsy10 scaled 1200}

\begin{document}
\title{A gaussian model of the dynamics of an inextensible chain}
\author{Franco Ferrari}
\email{ferrari@fermi.fiz.univ.szczecin.pl}
\author{Maciej Pyrka}\email{pyrolus@o2.pl}
\affiliation{Institute of Physics and CASA*, University of Szczecin,
  ul. Wielkopolska 15, 70-451 Szczecin, Poland}

\begin{abstract}
In this work an approximated path integral model describing the
dynamics of a 
inextensible chain is presented. To this purpose, the nonlinear
constraints which enforce the property of inextensibility of the chain
are relaxed and are just imposed in an average sense.
This strategy, which has been originally proposed for
semi-flexible polymers in statistical mechanics, is complicated in the
case of dynamics by the extra dependence on the time variable
and by the presence of nontrivial boundary conditions.
Despite these complications, the probability function of the chain,
which measures the probability to pass to a given initial conformation
to a final one, is computed exactly. The Lagrange multiplier
imposing the relaxed condition satisfies a complicated nonlinear
equation, which has been solved assuming that
the chain is very long.
\end{abstract}
\maketitle
\section{Introduction}\label{sec:intro}
The Rouse's \cite{rouse} and Zimm's \cite{zimm} models are able to
describe satisfactorily the 
dynamical behavior of a polymer in a solution \cite{doiedwards}. There
are however 
physical situations, such as for instance the case of a chain pulled
by strong forces in experiments of micromanipulation of DNA
\cite{bustamante,FEMIROVI,marko1,nelson, marko2}, whose 
correct interpretation requires that the chain is inextensible.
To implement the property of inextensibility, it is necessary
to introduce constraints in the
stochastic equations that govern its dynamics.
To this purpose, one may use generalized coordinates
or Lagrange multipliers. Alternatively, it
is possible to add forces, which become strongly repulsive whenever the
chain attempts to abandon the region of the phase space in which the
constraints are satisfied. Applying these methods one arrives
at equations which determine the dynamics of a chain in
the presence of constraints in a very rigorous way, but
are also very
complicated. Their solution, as much as the
practical calculation of physical
quantities, require numerical simulations. An
approach which leads to a simpler formulation of the problem consists
in considering the chain as a system of particles held together by an
elastic potential.
This potential must be chosen in such a way that in the equilibrium
position the distance between two neighboring particles in the chain
is equal to some constant $a\ne 0$. In  the limit of infinite elastic
constant, one obtains a discrete chain composed by particles connected
together 
by massless segments of length $a$. In the
limit in which $a$ goes to zero, while the number of particles becomes
infinite, one obtains the probability function of a continuous chain,
which measures the probability that the chain passes after a fixed
time $t_f$  from a given initial conformation to
a given final conformation. It is possible to show that such
probability function is equivalent to the
partition function of a nonlinear field theory, which has been
discussed in Ref.~\cite{FePaVi}. The appearance of
 nonlinearity is related to the
inextensibility constraints which require that at any
instant $t$ and at each point of the
chain, whose position is provided by the radius vector $\boldsymbol
R(t,s)$, the relation $\left|
\frac{\partial\boldsymbol R(t,s)}{\partial s}
\right|^2=1$ is satisfied. Here we have denoted with $s$ the
arc--length of the curve describing the spatial conformation of the
chain. The nonlinear model of Ref.~\cite{FePaVi} has a relatively
simple formulation in terms of path integrals and allows analytic
calculations, such as for example that of the dynamical form factor in
the semiclassical approximation \cite{FePaVi2}. Yet, 
it is important to have also a suitable approximation of that
model which is able to simplify the functional Dirac delta function
$\delta\left(
\left|
\frac{\partial\boldsymbol R(t,s)}{\partial s}
\right|^2-1
\right)$. 
This delta function appears in the model as a result of the
inextensibility constraints and it
 is a very nonlinear term. A similar delta function
has been simplified so far in
the case of the statistical mechanics of a freely hinged chain, where
the time variable is not present, using the substitution
$ \delta\left(
\left|
\frac{\partial\boldsymbol R(s)}{\partial s}
\right|^2-1
\right)\longrightarrow
e^{\frac{3}{2a}\int_0^Lds\left(\frac{\partial\boldsymbol
    R(s)}{\partial s} \right)^2
}$ \cite{EdwGoo}. 
In the latter formula, $L$ represents the total length of the chain.
Unfortunately, the extension of the approach of Ref.~\cite{EdwGoo} to
polymer dynamics is not feasible.
To obtain a Gaussian approximation of the functional delta function
comparable to  that proposed by the authors of \cite{EdwGoo} also in the
case of dynamics, we use here an alternative strategy borrowed from the
statistical mechanics of semi-flexible
chains\cite{thirumalai1,thirumalai2}. The 
idea is to loose the 
condition that all arcs composing the chain must be inextensible
and to require  instead
that the time averaged total length of the chain
should fluctuate around a fixed average value $L$.
This means that the length of the chain is allowed to change in time
as it happens in the Rouse model, but only in such a way that its
average is equal to $L$.
This relaxed constraint is imposed in the probability function of
the chain by introducing
a Lagrange multiplier $\lambda$, which is
determined by minimizing
the partition function as a function of $\lambda$. With respect to the
case of semi-flexible polymers, the situation is complicated by
nontrivial boundary conditions and the presence
of the time.
At the end we are able to compute exactly
 the explicit expression of the probability function, but 
for its minimization it is necessary to solve a
complicated algebraic equation in
 $\lambda$. 
We find its solution in
the limit of very large values of the average total length of the
chain $L$ up to corrections of the order $\frac 1L$ included.
\section{The gaussian approximation}
We consider in this work the partition function of the model
\cite{FePaVi}:
\beq{Z=\int_{\boldsymbol R(0,s)=\boldsymbol R_0(s)}^{\boldsymbol R(t_f,s)=\boldsymbol R_f(s)}
{\cal D}\boldsymbol
R(t,s)e^{-c\int_0^{t_f}dt\int_0^Lds\dot{\boldsymbol
    R}^2(t,s)}\delta(\boldsymbol R'^2(t,s)-1)
}{action}
where $\boldsymbol R(t,s):[0,t_f]\times[0,L]\longrightarrow \mathbb
{R}^d $ is a two dimensional vector field with $d$ components.
The partial derivatives of $\boldsymbol R(t,s)$ with respect to $t$
and $s$ are denoted as follows:
\beq{
\dot{\boldsymbol R}(t,s)=\frac{\partial \boldsymbol R(t,s)}{\partial
  t}
\qquad\qquad
{\boldsymbol R}'(t,s)=\frac{\partial \boldsymbol R(t,s)}{\partial s}
}{notationone}
The parameter $c$ appearing in the action \ceq{action} is given by:
\beq{
c=\frac{M}{2L}\frac 1{2k_BT\tau}
}{parc}
$k_B$ denotes the Boltzmann  constant, $T$ is  the fixed temperature
of the  thermal bath in which the chain fluctuates and $\tau$ is  the
relaxation time which characterizes the rate with which the
infinitesimal beads lose their speed due to friction. Finally, $M$ is
the total mass of the chain.

It has been shown in \cite{FePaVi,FePaJPA} that this model
describes the dynamics of an inextensible continuous chain of length
$L$ obtained by performing the continuous limit of the partition
function of a  freely jointed chain. The dynamics of the chain
is followed during the period of time $t\in[0,t_f]$.
$s\in[0,L]$ is the arc-length
measuring the distance along the chain.
The vector field $\boldsymbol R(t,s)$ denotes the positions of the
infinitesimal elements of length $ds$ composing the chain at a given
instant $t$.
$Z$ has the meaning of the probability function which measures the
probability that the chain during its fluctuations passes from the
initial conformation $\boldsymbol R_0(s)$ at the instant $t=0$ to the
final conformation $\boldsymbol R_f(s)$ at the instant $t=t_f$.
The constraints enforcing the conditions that the length of the
links connecting the beads should be constant become in the
continuous limit the constraint:
\beq{
\boldsymbol{R}^{\prime 2}(t,s)=1
}{constr}
which is imposed in Eq.~\ceq{action} using a functional Dirac delta function.
It is easy to check that this constraint implies that the length of
every part of 
the chain is constant in time. Let us take for instance an arc of the
curve $\boldsymbol R(t,s)$ delimited by the points $\boldsymbol
R(t,s_1)$
and $\boldsymbol R(t,s_2)$. Its length $\ell_{12}(t)$ is given by:
$\ell_{12}(t)=\int_{s_1}^{s_2}ds|\boldsymbol R'(t,s)|$. Due to
Eq.~\ceq{constr} it is possible to write:
$\ell_{12}(t)=\int_{s_1}^{s_2}ds=s_2-s_1$. Thus 
\beq{\dot{\ell}_{12}(t)=0}{constlengthpart}

The probability function $Z$ in Eq.~\ceq{action} should be completed by
appropriated boundary conditions. We require that at the initial and
final 
instants $0$ and $t_f$ the chain is in  the configurations
$\boldsymbol R_0(s)$ and $\boldsymbol R_f(s)$ respectively, i.~e.:
\begin{eqnarray}
\boldsymbol R(t_f,s)&=&\boldsymbol R_f(s)\label{bcone}\\
\boldsymbol R(0,s)&=&\boldsymbol R_0(s)\label{bctwo}
\end{eqnarray}
Moreover, we suppose that the chain is closed, so that periodic
boundary conditions will be chosen with respect to $s$:
\beq{
\boldsymbol R(t,s)=\boldsymbol R(t,s+L)
}{boundconds}
In the following, it will be convenient to expand the boundary
conformations in Fourier series:
\begin{eqnarray}
\boldsymbol R_f(s)&=&\sum_{n=-\infty}^{+\infty}e^{2\pi i n\frac
  sL}\boldsymbol b_n\\
\boldsymbol R_0(s)&=&\sum_{n=-\infty}^{+\infty}e^{2\pi i n\frac
  sL}\boldsymbol a_n
\end{eqnarray}
$\boldsymbol a_n$ and $\boldsymbol b_n$ being constant vectors.
The partition function of the model described by Eq.~\ceq{action} has
been computed in \cite{FePaVi} in the semiclassical approximation. It
is also possible to derive an expression of the dynamic structure
factor of the chain always in the semiclassical approximation as shown
in \cite{FePaVi2}. Despite these successes, the presence of the
functional Dirac delta function in the right hand side of
Eq.~\ceq{action} makes the analytical treatment of the model
complicated. For this reason, it would be nice to simplify it with a
suitable 
approximation. For instance, in the case of statistical mechanics of
polymers, where the time 
variable is not appearing, Edwards and Goodyear \cite{EdwGoo} have
shown for $d=3$ that:
\beq{
\delta(\boldsymbol R^{\prime 2}(s)-a^2)\sim e^{-\int_0^Lds\frac
  3{2a}\boldsymbol R^{\prime2}(s)}
}{statapprox}
While the extension of the results of \cite{EdwGoo} to dynamics is not
straightforward, it is however possible to relax the constraint
\ceq{constr} requiring that it is satisfied only in an average
sense. To this purpose, following a  procedure borrowed from
the statistical mechanics of semi-flexible polymers, see for instance
\cite{thirumalai1} 
and \cite{thirumalai2,winkler}, we 
introduce a new parameter $\lambda$ and replace the probability function
\ceq{action} with the following simplified one:
\beq{
Z(\lambda)=\int_{\boldsymbol R(0,s)=\boldsymbol R_0(s)}^{\boldsymbol
  R(t_f,s)=\boldsymbol R_f(s)} 
{\cal D}\boldsymbol R(t,s) e^{-\int_0^{t_f}dt\int_0^Lds\left[
c\dot{\boldsymbol R}^2(t,s)+\lambda(\boldsymbol R^{\prime 2}(t,s)-1)
\right]}
}{zlambda}
Next, the parameter $\lambda$ is determined by requiring that:
\beq{
\frac{\partial Z(\lambda)}{\partial\lambda}=0
}{condavefirst}
Clearly, the above equation is equivalent to:
\beq{
\left\langle
\frac 1{t_f}\int_0^{t_f}dt\frac 1L\int_0^Lds \boldsymbol R^{\prime 2}(t,s)
\right\rangle=1
}{relconstr}
where
\beq{
\left\langle(\ldots)\right\rangle=
\int_{\boldsymbol R(0,s)=\boldsymbol R_0(s)}^{\boldsymbol
  R(t_f,s)=\boldsymbol R_f(s)} 
{\cal D}\boldsymbol R(t,s) e^{-\int_0^{t_f}dt\int_0^Lds\left[
c\dot{\boldsymbol R}^2(t,s)+\lambda(\boldsymbol R^{\prime 2}(t,s)-1)
\right]}(\ldots)
}{fdfsdf}
In words, with \ceq{relconstr} the functional integration 
over the conformations  $\boldsymbol R(t,s)$ 
has been extended to curves
whose arcs do no longer satisfy the
constant length condition of Eq.~\ceq{constlengthpart}.
Also the total length of the chain is allowed to change in
time, but its
average  performed over all possible
conformations and during the period of time $t_f$ must be equal  to $L$.
\section{Calculation of the probability function $Z(\lambda)$}
The rest of this work is dedicated to the derivation of the
probability function $Z(\lambda)$ and of the parameter
$\lambda$, which up to now is the still unknown ingredient of
Eq.~\ceq{zlambda}. 
Unlike the case of semi-flexible polymers, we have to deal here with
the nontrivial boundary conditions \ceq{bcone} and \ceq{bctwo}.
To get rid of them, it is convenient to split the bond vectors
$\boldsymbol R(t,s)$  into two components:
\beq{
\boldsymbol R(t,s)=\boldsymbol R_{cl}(t,s)+\boldsymbol r(t,s)
}{splitting}
$\boldsymbol R_{cl}(t,s)$ is a solutions of the classical equations of
motion related to the action:
\beq{
S=\int_0^{t_f}dt\int_0^Lds\left[
c\dot{\boldsymbol R}^2(t,s)+\lambda(\boldsymbol R^{\prime 2}-1)
\right]
}{effaction}
appearing in the probability function of Eq.~\ceq{zlambda}:
\beq{
c\ddot{\boldsymbol R}_{cl}+\lambda\boldsymbol R^{\prime\prime}_{cl}=0
}{clasequmot}
$\boldsymbol R_{cl}(t,s)$ satisfies the boundary conditions
\ceq{bcone} and \ceq{bctwo}:
\beq{
\boldsymbol R_{cl}(t_f,s)=\boldsymbol R_f(s)\qquad\qquad
\boldsymbol R_{cl}(0,s)=\boldsymbol R_0(s)
}{bdcondd}
$\boldsymbol r(t,s)$ represents instead the fluctuations around the
classical conformation $\boldsymbol R_{cl}(t,s)$. $\boldsymbol r(t,s)$
satisfies the Dirichlet boundary conditions:
\beq{
\boldsymbol r(t_f,s)=\boldsymbol r(0,s)=0
}{dirboucon}
With respect to the variable $s$ both $\boldsymbol R_{cl}(t,s)$ and
$\boldsymbol r(t,s)$ satisfy periodic boundary conditions analogous to
those of Eq.~\ceq{boundconds}.
After some calculations and without performing any approximation, it
is possible to show that:
\beq{
Z(\lambda)=e^{-S(\boldsymbol R_{cl})}e^{Lt_f\lambda}Z_{fluct}(\lambda)
}{zlambaint}
where the action $S(\boldsymbol R_{cl})$ contains the contribution
coming from the classical background conformations:
\beq{
S(\boldsymbol R_{cl})=\int_0^{t_f}dt\int_0^Lds\left(
c\dot{\boldsymbol R}^2_{cl}+\lambda\boldsymbol R^{\prime 2}_{cl}
\right)
}{contrclaback}
while $Z_{fluct}(\lambda)$ is the partition function of the
fluctuations:
\beq{
Z_{fluct}(\lambda)=\int{\cal D}\boldsymbol r
e^{-\int_0^{t_f}dt\int_0^Lds\left(
c\dot{\boldsymbol r}^2+\lambda\boldsymbol r^{\prime 2}
\right)}
}{zfluct}
First, we will compute $S(\boldsymbol R_{cl})$.
It is possible to check after some integrations by parts that
\beq{
S(\boldsymbol R_{cl})=c\int_0^Lds\left(
\boldsymbol R_f(s)\cdot\dot{\boldsymbol R}_{cl}(t_f,s)-\boldsymbol
R_0(s)\cdot \dot{\boldsymbol R}_{cl}(0,s)
\right)
}{srcla}
with $\dot{\boldsymbol R}_{cl}(t_f,s)=\left.\frac{\partial \boldsymbol
  R_{cl}(t,s)}{\partial t}\right|_{t=t_f}$ and
$\dot{\boldsymbol R}_{cl}(0,s)=\left.\frac{\partial \boldsymbol
  R_{cl}(t,s)}{\partial t}\right|_{t=0}$.
The solution of the classical equations of motion \ceq{clasequmot}
satisfying the boundary conditions of Eq.~\ceq{bdcondd} is:
\beq{
\boldsymbol R_{cl}(t,s)=\sum_{n=-\infty}^{+\infty}e^{2\pi in\frac
  sL}\left(
-\frac{\boldsymbol a_n}{\sinh (\beta_nt_f)}\sinh(\beta_n(t-t_f))+
\frac{\boldsymbol b_n}{\sinh (\beta_nt_f)}\sinh(\beta_nt)
\right)
}{rcla}
where
\beq{
\beta_n=\sqrt{\frac \lambda c}\frac {2\pi n}L
}{betandef}
Substituting the expression of the classical solution $\boldsymbol
R_{cl}$ of Eq.~\ceq{rcla} in Eq.~\ceq{srcla}, we obtain:
\beq{
S(\boldsymbol
R_{cl})=-cL\sum_{n=-\infty}^{+\infty}\beta_{-n}\left[-
\frac 1{\sinh(\beta_{-n}t_f)}\left( 
\boldsymbol b_n\cdot\boldsymbol a_{-n}+\boldsymbol a_n\cdot\boldsymbol b_{-n}
\right) 
+ \frac{\cosh(\beta_{-n}t_f)}{\sinh(\beta_{-n}t_f) }
\left( 
\boldsymbol b_n\cdot\boldsymbol b_{-n}+\boldsymbol a_n\cdot\boldsymbol a_{-n}
\right) 
\right]
}{srclcom}
Next, we derive the partition function of the fluctuations
$Z_{fluct}(\lambda)$. From Eq.~\ceq{zfluct} it turns out in $d=3$ that:
\beq{
Z_{fluct}(\lambda)=({\det}'\Delta)^{-\frac 32}
}{detfluct}
where $\Delta$ is the differential operator:
\beq{
\Delta=c\frac{\partial^2}{\partial t^2}+\lambda
\frac{\partial^2}{\partial s^2} 
}{deltadiffop}
The prime after the symbol of determinant in Eq.~\ceq{detfluct} means
that the zero modes of the operator $\Delta$ should be projected out
from the expression of the determinant.
At this point we compute the eigenvalues of $\Delta$. To this purpose,
we have to solve the equation:
\beq{
\left(
c\frac{\partial^2}{\partial t^2}+\lambda \frac{\partial^2}{\partial s^2}
\right)\psi=-E\psi
}{eigprob}
Dirichlet boundary conditions with respect to the time variable are
assumed according to Eq.~\ceq{dirboucon}. Moreover, $\psi$ should
obey periodic boundary conditions with respect to $s$.
It is possible to check that there exist solutions of
Eq.~\ceq{eigprob} with such boundary conditions when $E$ takes the
values:
\beq{
E_{n,k}=4\pi^2\left(
n^2\frac\lambda{L^2}+k^2\frac c{2t_f^2}
\right)
\qquad\qquad n,k=0,\pm1,\pm2,\ldots
}{eigenvalues}
These solutions are the eigenfunctions:
\beq{
\psi_{n,k}(t,s)=\sum_{n=-\infty}^{+\infty}e^{2\pi i n\frac sL}\psi_{n,k}(t)
}{eigenfunctions}
where
\beq{
\psi_{n,k}(t)=A\sinh\left[
t\sqrt{\frac 1c\left(
\lambda\frac {4\pi^2n^2}{L^2}+E_{n,k}
\right)}
\right]
}{eigenfunctionst}
and $A$ is a normalization constant.
Excluding the trivial zero mode occurring for $n=k=0$, we find that
\beq{
{\det}^\prime\Delta={\textstyle\prod_{n,k\ge 0}^{\prime}}\left[4\pi^2\left(
n^2\frac\lambda{L^2}+k^2\frac c{2t_f^2}
\right)\right]
}{detprime}
In the above equation the prime in the product $\prod_{n,k\ge 0}'$
means that the case  $n=k=0$, which corresponds to the excluded zero
mode, should be omitted. 
Now we rewrite ${\det}^\prime\Delta $ as follows:
\beq{
{\det}^\prime\Delta=\left(
{\textstyle\prod_{n,k\ge 0}'}\frac\lambda{L^2}
\right)
{\textstyle\prod_{n,k\ge 0}^{\prime}}\left[4\pi^2\left(
n^2+\frac{cL^2k^2}{2t_f^2\lambda}
\right)\right]
}{deltaprimeaa}
Using the $\zeta-$function regularization it is possible to show that
${\textstyle\prod_{n,k\ge 0}'}\frac\lambda{L^2}=\left(
\frac \lambda{L^2}
\right)^{\frac 54}$. Thus:
\beq{
{\det}^\prime\Delta=\left(
\frac \lambda{L^2}
\right)^{\frac 54}{\textstyle\prod_{n,k\ge 0}^{\prime}}\left[4\pi^2\left(
n^2+\frac{cL^2k^2}{2t_f^2\lambda}
\right)\right]
}{deltaprimebb} 
The remaining semi-infinite product has been evaluated by several
authors, see for instance
\cite{semiinf1,semiinf2,semiinf3,meissner,meissner2}. Here we present
just the result of the calculations:
\beq{
{\textstyle\prod_{n,k\ge 0}^{\prime}}\left[4\pi^2\left(
n^2+\frac{cL^2k^2}{2t_f^2\lambda}
\right)\right]=\exp\left(
\ln\tau-\frac{\pi\tau}{12}+\sum_{n>0}\ln\left(
1-e^{-2\pi n\tau}
\right)
\right)
}{deltaprimecc}
where
\beq{\tau^2=\frac{2t_f^2\lambda}{cL^2}}{taudef}
Summarizing, the partition function of the fluctuations given in
Eq.~\ceq{detfluct} becomes:
\beq{
Z_{fluct}(\lambda)=\exp\left[
-\frac {15}{8}
\ln\left(\frac{\tau^2c}{2t_f^2}\right)-\frac 32\ln\tau+\frac{\pi\tau}{8}
-\frac32\sum_{n>0}\ln\left(
1-e^{-2\pi n\tau}
\right)\right]
}{zfluctfinal}
Putting together the results of Eqs.~\ceq{srclcom} and
\ceq{zfluctfinal}, the total probability function $Z(\lambda)$ of
Eq.~\ceq{zlambaint} may be rewritten as follows:
\beq{
Z(\lambda)=e^{F(\lambda)}
}{zlambaintint}
where
\beq{
F(\lambda)=-S(\boldsymbol R_{cl})+Lt_f\lambda -\frac{15}8\ln\left(
\frac{\tau^2c}{2t_f^2}\right)
-\frac 32\ln\tau+\frac{\pi\tau}8-\frac
32\sum_{n>0}\ln \left(
1-e^{-2\pi\tau}
\right)
}{dasd}
Clearly, the condition \ceq{condavefirst} that determines $\lambda$ is
equivalent to the condition $\frac{\partial F(\lambda)}{\partial
  \lambda}=0$, i.~e.:
\beq{
\frac{\partial S(\boldsymbol R_{cl})}
{\partial \lambda}
-Lt_f
+
\frac{21}8
\frac 1\lambda 
-\frac\pi 8
\frac {t_f}L
\sqrt{
\frac 1{2c}}
\frac 1{\sqrt{\lambda}}
+\frac 32\sum_{n>0}
\frac{2\pi n
}{1-e^{-2\pi n \tau}}e^{-2\pi n\tau}
\frac {t_f}L
\sqrt{\frac 1{2c}}\frac 1{\sqrt{\lambda}}=0
}{conda}
As it stands, the above equation is too difficult to be solved
analytically with respect to $\lambda$. For this reason, we assume
that the chain is very long, i.e. $L>>1$, so that it is possible to
expand $\lambda$ and the coefficients appearing in
Eq.~\ceq{conda} in powers of $L^{-1}$.
In doing that, we suppose that $\lambda$ remains finite in the limit
$L\longrightarrow+\infty$, so that:
\beq{
\lambda\sim\lambda_0+\frac 1L\lambda_1+\ldots
}{lambdaexp}
This hypothesis on $\lambda$, which
will be checked a posteriori, has a physical motivation.
In fact, if $\lambda$ becomes
infinite with the average chain length $L$, then  conformations for which
$\frac 1{t_f}\int_0^{t_f}dt\frac 1L\int_0^Lds\boldsymbol{R}^{\prime
  2}<1$ will be strongly preferred in the probability function of
Eq.~\ceq{zlambda}, a fact which is clearly 
unphysical. 

At this point, we are ready to expand all the terms
entering in Eq.~\ceq{conda} in powers of $\frac 1L$.
This expansion is almost trivial apart from the case of the
two terms
\beq{
I_1=\frac{\partial S(\boldsymbol
  R_{cl})}{\partial \lambda}}{Ione} and 
\beq{I_2=\frac 32\sum_{n>0}
\frac{2\pi n
}{1-e^{-2\pi n \tau}}e^{-2\pi n\tau}
\frac {t_f}L
\sqrt{\frac 1{2c}}\frac 1{\sqrt{\lambda}}}{Itwo}
$I_1$ depends on $\lambda$ implicitly through the
coefficients $\beta_{n}$,
 while $I_2$ has an explicit dependence
on $\lambda$ together with an implicit dependence through the
coefficient $\tau$.
Provided Eq.~\ceq{lambdaexp} is
fulfilled \cite{note1},
we note that 
both parameters $\beta_n$ and $\tau$, being proportional to $\frac 1L$,
are going to zero when $L$ goes to infinity.
Using the above considerations and neglecting contributions of order
$\frac 1L$ or higher, we obtain: 
\begin{eqnarray}I_1&\sim&
  0\label{approxterm1} \\
I_2&\sim&\frac 3{8\pi}\frac
L{t_f}\sqrt{\frac c2} \frac 1{\lambda_0^{\frac 32}}-\frac
3{4\lambda_0}-\frac 9{16}\frac 1{\pi t_f}\sqrt{\frac c2}\frac
{\lambda_1}{\lambda_0^{\frac 52}}
\label{approxterm2}
\end{eqnarray}
After completing the expansion in powers of $\frac 1L$ of the
remaining terms, Eq.~\ceq{conda} becomes:
\beq{
L\left(
\frac {3}{8\pi t_f}\sqrt{\frac c2}\frac 1{\lambda_0^{\frac 32}}-t_f
\right)+\left(
\frac{15}{8}\frac 1{\lambda_0}-\frac 9{16\pi t_f}\sqrt{\frac
  c2}\frac{\lambda_1}{\lambda_0^{\frac 52}}
\right)+{\cal O}(\frac 1L)=0
}{condaapprox}
Equating to zero separately the coefficients accompanying different
powers of $L$ we obtain two relations which are able to determine both
$\lambda_0$ and $\lambda_1$:
\beq{
\lambda_0=\sqrt[3]{\frac {9c}{128\pi^2t_f^4}}
}{dffdzero}
\beq{\lambda_1=\frac 5{4t_f}
}{dffdone}
As a result it is possible to write down the
expansion of $\lambda$ which is valid for large values of $L$ up to
the order $\frac 1L$:
\beq{
\lambda\sim\sqrt[3]{\frac {9c}{128\pi^2t_f^4}}+\frac 1L\frac 5{4t_f}
}{final}
This, together with Eqs.~(\ref{zlambaintint}--\ref{dasd}) concludes
the evaluation of the probability 
function \ceq{zlambda}. 
\section{Conclusions}
In this article a gaussian approximation of the nonlinear model
\ceq{action} has been presented, which describes the probability of a
chain to pass from a given conformation to another. The approximation
consists in replacing the local constraint \ceq{constr} with a milder
one, that requires only that during the fluctuations of the chain, its
average 
length must be equal to $L$. To enforce this new
constraint, the Lagrange multiplier $\lambda$ has been
introduced. $\lambda$ has been determined exploiting the condition
that, as a function of $\lambda$, the probability function
$Z(\lambda)$ of Eq.~\ceq{zlambda} has a minimum. Eq.~\ceq{relconstr}
shows that this condition is equivalent to impose the relaxed
constraint. While it is possible to compute the probability function
$Z(\lambda)$ exactly despite the complications due to the presence of
nontrivial boundary conditions, its minimization requires the solution
with respect to $\lambda$ 
of the complicated algebraic equation given
in Eq.~\ceq{conda}. This equation has been solved in the limit of very
large values of $L$ up to the second order included, see
Eq.~\ceq{final}.
It is shown that $\lambda$ goes to zero for increasing values of the
evolution time $t_f$. This means that the deviations of the length of
the chain from the expected value $L$ increase with the increase of
the time in which the chain is allowed to fluctuate.

The present approach avoids the difficulties involved with the
functional delta function of the original probability function
\ceq{action} and it is still able to take into account the
inextensibility of the chain, at least in an approximated way. The
probability function $Z(\lambda)$ of Eq.~\ceq{zlambda} with the
relaxed constraint has been computed here exactly, see
Eqs.~(\ref{zlambaintint}--\ref{dasd}), while the analogous
probability function with the full constraint \ceq{action} could be
derived only in semiclassical approximation.
The approximated model of Eq.~\ceq{zlambda} may be used as a basis for
further computations of 
physical observables, such as for instance the dynamic form
factor of a chain.
\section{Acknowledgments}
The authors wish to thank J. Paturej for  fruitful discussions and
useful remarks.
One of the author (F. F.) would also like to thank
Z. Jask\'olski, V. G. Rostiashvili and T. A. Vilgis for
their remarks and suggestions. 
M. Pyrka would also like to thank the Institute of Physics and the Faculty
of Mathematics and Physics of the University of Szczecin for their
hospitality.


\begin{thebibliography}{99}
\bibitem{rouse} P. E. Rouse, {\it J. Chem.
    Phys.} {\bf 21} (1953), 1272.  
\bibitem{zimm} B. H. Zimm, {\it J. Chem.
    Phys.} {\bf 24} (1956), 269.  
\bibitem{doiedwards} M. Doi and S.F. Edwards, The Theory of Polymer Dynamics
  (Clarendon Press, Oxford, 1986).  
\bibitem{bustamante} C. Bustamante et Al., {\it Current Opinion in
  Structural Biology} {\bf 10} (3) (2000), 279.
\bibitem{FEMIROVI} M. Febbo, A. Milchev, V. Rostiashvili,
  T. A. Vilgis, and D. Dimitrov, 
{\it J. Chem. Phys} {\bf 129} (2008), 154808.
 arXiv:0808.0891.
\bibitem{marko1} C. Bustamante, J. F. Marko, E. D. Siggia and
  S. Smith, {\it Science} {\bf 265} (1994), 415.
\bibitem{nelson} C. Storm and P. C. Nelson,
{\it Phys. Rev.} {\bf E 67} (2003),  051906.
\bibitem{marko2} J. F. Marko and E. D. Siggia, {\it Macrom.} {\bf 28}
  (1995),
 8759.
\bibitem{FePaVi} F. Ferrari, J. Paturej and T. A. Vilgis, {\it
  Phys. Rev. E}, {\bf 77},  021802, 
  2008.
\bibitem{FePaVi2} F. Ferrari, J. Paturej and T. A. Vilgis, 
{\it Phys. Atomic Nuclei} ({\it Journal of Yadernaya Fizika}) {\bf 73}
(2) (2010), 316. 
\bibitem{EdwGoo} S. F. Edwards and A. G. Goodyear, {\it J. Phys. A:
  Gen. Phys.} {\bf 5} (1972), 965.
\bibitem{FePaJPA}
F. Ferrari and J. Paturej,
{\it Jour. Phys. A: Mathematical and Theoretical}, {\bf 42} (14)
(2009), 145002.
\bibitem{thirumalai1} B.-Y. Ha and D. Thirumalai, {\it J. Chem. Phys.}
  {\bf 103} (1995), 9408.
\bibitem{thirumalai2} B.-Y. Ha and D. Thirumalai, {\it J. Chem. Phys.}
 {\bf 106} (1997), 4243. 
\bibitem{winkler} R. G. Winkler, {\it J. Chem. Phys.} {\bf 118} (6)
  (2003), 2919.
\bibitem{semiinf1} O. Alvarez, {\it Nucl. Phys.} {\bf B216} (1983),
  125.
\bibitem{semiinf2} J. Polchinski, {\it Comm. Math. Phys.} {\bf 104}
  (1986), 37.
\bibitem{semiinf3} G. Moore and P. Nelson, {\it Nucl. Phys.} {\bf B266}
  (1986), 58.
\bibitem{meissner} K. A. Meissner and S. Pokorski, {\it Z. Phys. C}
  {\bf 36} (1987), 105.
\bibitem{meissner2} K. A. Meissner and S. Pokorski, {\it Acta
  Phys. Pol.} {\bf B19} (1988), 659.
\bibitem{note1} More in general
  provided that $\lambda$ does not grow with increasing values of $L$
  with power law $\lambda\propto L^{2+\epsilon}$ with $\epsilon\ge0$.
\end{thebibliography}
\end{document}